\documentclass{elsart}

\usepackage{amsmath,amssymb,graphicx,bm}
\usepackage{color}

\newcommand{\beq}{\begin{equation}}
\newcommand{\eeq}{\end{equation}}
\newcommand{\bea}{\begin{eqnarray}}
\newcommand{\eea}{\end{eqnarray}}
\newcommand{\ba}{\begin{align}}
\newcommand{\ea}{\end{align}}
\newcommand{\bfig}{\begin{figure}}
\newcommand{\efig}{\end{figure}}

\newcommand{\lam}{\lambda}
\newcommand{\la}{\Lambda}

\newcommand{\vlowk}{V_{{\rm low}\,k}}

\newcommand{\kf}{k_{\text{F}}}

\newcommand{\fmi}{\, \text{fm}^{-1}}
\newcommand{\mev}{\, \text{MeV}}

\newcommand{\ef}{\mu}
\newcommand{\ek}{\varepsilon_{\rm k}}

\newcommand{\Gmed}{ G^{0}_{pphh}}

\newcommand{\Gprop}{{\cal G}}
\newcommand{\Gtildeprop}{\widetilde{\cal G}}
\newcommand{\Fprop}{{\cal F}}
\newcommand{\Fdagprop}{{\cal F}^{\dagger}}

\newcommand{\veck}{\vec{k}}

\newcommand{\Ecm}{E}

\newcommand{\kprime}{k^{\prime}}

\newcommand{\Hzero}{H_0}
\newcommand{\etaNG}{\eta_{\nu}^{\rm NG}}
\newcommand{\D}{\displaystyle}

\newcommand{\lan}{\langle}
\newcommand{\ran}{\rangle}

\begin{document}

\begin{frontmatter}

\title{Weinberg Eigenvalues and Pairing \\ with Low-Momentum Potentials}
\author[OSU,CHEP]{S.\ Ramanan}
\ead{suna@pacific.ohio-state.edu},
\author[OSU,MSU]{S.K.\ Bogner},
\ead{bogner@mps.ohio-state.edu}
and
\author[OSU]{R.J.\ Furnstahl}
\ead{furnstahl.1@osu.edu}

\address[OSU]{Department of Physics,
The Ohio State University, Columbus, OH\ 43210}
\address[CHEP]{Center for High Energy Physics,
Indian Institute of Science, Bangalore\ 560012}
\address[MSU]{National Superconducting Cyclotron Laboratory and Department of 
Physics and Astronomy, Michigan State University, East Lansing, MI 48844}

\date{\today}

\begin{abstract}
%
The nonperturbative nature of nucleon-nucleon interactions
evolved to low momentum has recently been investigated in
free space and at finite density using Weinberg eigenvalues as a diagnostic.
This analysis is extended here to the in-medium eigenvalues near the Fermi
surface to study pairing. For a fixed value of density and cutoff, 
the eigenvalues increase
arbitrarily in magnitude close to the Fermi surface, signaling the pairing 
instability. 
When using normal-phase propagators, 
the Weinberg analysis with complex energies
becomes a form of stability analysis and 
the pairing gap can 
be estimated from the largest attractive eigenvalue. 
With Nambu-Gorkov Green's functions, the largest attractive eigenvalue 
goes to unity close to the Fermi surface, 
indicating the presence of bound states 
(Cooper pairs), and the corresponding 
eigenvector leads to the self-consistent gap function. 
\end{abstract}

\end{frontmatter}

\section{Introduction}
\label{sect:intro}

The perturbativeness of a nucleon-nucleon (NN) potential  can be
quantified using the eigenvalue analysis introduced long ago by
Weinberg \cite{Weinberg}.
Consider the operator Born series for the free-space $T$-matrix at energy $E$:
\beq
 T(E) = V + V \frac{1}{E-\Hzero} V + \cdots
\eeq 
By finding the eigenvalues and eigenvectors of 
\beq
  \frac{1}{E - \Hzero} V | \Psi_\nu \rangle
    = \eta_\nu(E) | \Psi_\nu \rangle
    \; ,
    \label{eqn:free_space}
\eeq
and then acting with $T(E)$ on the eigenvectors,
\beq
 T(E) | \Psi_\nu \rangle
   =  V | \Psi_\nu \rangle
   (1 + \eta_\nu + \eta_\nu^2 + \cdots)
   \; ,
\eeq
it follows that nonperturbative behavior at energy $E$ is signaled
by one or more eigenvalues with 
$|\eta_\nu(E)| \geq 1$ \cite{Weinberg,fewbody}.
Such an analysis has been recently used as a diagnostic of 
low-momentum potentials in free 
space~\cite{Bogner:2006tw,Bogner:2006vp,Bogner:2006pc}.
Major decreases in the magnitudes of the largest eigenvalues were
observed 
as bare nucleon-nucleon potentials (such as those
from Refs.~\cite{AV18,N3LO}) 
were evolved using  
renormalization group (RG) methods~\cite{VlowkRG,VLOWK}. These decreases can 
be linked to a dampening of the sources of nonperturbative 
physics, such as the repulsive core and the short-range tensor
interaction, as the cutoff is lowered. 
Lowering $\la$ yields a soft potential
(generically called ``$\vlowk$''), which in turn simplifies few and many-body
calculations~\cite{perturbative,Bogner:2005fn}.\footnote{There are
various RG methods used to generate low-momentum potentials, including
using sharp and smooth regulators and 
through the Similarity Renormalization Group
(SRG)~\cite{Bogner:2006pc}. SRG potentials depend on a parameter
$\lam$, which measures the spread of the off-diagonal strength and acts
as a cutoff. 
We restrict our calculations here to sharp-cutoff $\vlowk$ potentials;
gaps with smooth regulators are discussed in
Ref.~\cite{HebelerSchwenk:2006}.}       

The Weinberg analysis was extended to the in-medium T-matrix
in Ref.~\cite{perturbative} to examine the effect of Pauli blocking.
The conclusion was that the bulk nuclear matter energy calculations
are perturbative in the particle-particle channel.
The focus in Ref.~\cite{perturbative} was primarily on repulsive eigenvalues, 
which are tied to short-range physics.
But questions naturally arise about other sources of non-perturbative
physics.
In this paper we extend the in-medium eigenvalue analysis to energies
close to the 
Fermi surface, where the attractive NN interaction leads 
to a pairing instability in the particle-particle channel. 
This nonperturbative feature should be reflected in the Weinberg
eigenvalues. 
Verifying the eigenvalue analysis for pairing is also a step
toward a more general application of this tool to long-range
correlations in the particle-hole channel. 

In Sec.~\ref{sect:weinberg}, we review the in-medium results and
focus on the Weinberg eigenvalues for energies close to the Fermi surface.
This takes the form of a stability analysis with
complex energies~\cite{schrieffer}. From the eigenvalues it is possible to
estimate the BCS gaps. Pairing is
naturally taken into account in the Nambu-Gorkov formalism. 
Using the two-particle Nambu-Gorkov Green's
function we evaluate the Weinberg eigenvalues in Sec.~\ref{sect:nambu}, 
first for a separable model and
then for a $\vlowk$ potential. 
Throughout this paper we work at zero temperature, with
only a two-body potential and a free single-particle spectrum. Moreover we
consider only the $^1$S$_0$ partial wave, although the analysis can
be applied more generally and 
the cutoff dependence of the extracted gap used to test the importance of 
three-body forces in other partial-wave channels. We summarize our
findings and discuss future investigations in 
Sec.~\ref{sect:conclusion}.


\section{In-medium Weinberg Eigenvalues and the Pairing Gap}
 \label{sect:weinberg}

In free space, the Weinberg eigenvalue analysis refers to
the spectrum of $V G^0_{pp}$ as in 
Eq.~(\ref{eqn:free_space}), where
$G^0_{pp}$ is the two-body non-interacting particle-particle Green's function:
\beq
	G^0_{pp}(E) = \D\frac{1}{E - \Hzero} \;.
\eeq
As noted above, when the magnitude of the largest eigenvalue lies
outside the unit circle, i.e., $|\eta_{\nu}(E)| \ge 1$, the
corresponding Born series  expansion for the $T$ matrix no longer
converges. 
Lowering the cutoff results in a softening of the short-range
repulsion and the iterated tensor force in free space, while the impact
of shallow bound states in the S~waves 
is eliminated at sufficiently high density due to
Pauli-blocking.   As a result the eigenvalues decrease in magnitude as
the cutoff is lowered. In free space, for any value of the  cutoff
there are only a finite number of eigenvalues which lie outside the 
unit circle~\cite{Weinberg}. 

The same conclusions hold in-medium as long as we work at energies
away from the
Fermi-surface~\cite{Bogner:2006tw,perturbative}.
Figure~\ref{med.fig0}
shows the largest attractive and repulsive eigenvalues at center of mass energy $\Ecm = 0
 \mev$ for neutron matter in the $^1S_0$ channel for three different
values of the cutoff $\la$.%
\footnote{At negative energy, the purely real free-space Weinberg eigenvalues 
can
be viewed as inverse coupling constants that the interaction must be
scaled by to support a bound state at that energy. Therefore, negative and
positive Weinberg eigenvalues are called repulsive and attractive
eigenvalues.  The same designations are used for positive-energy
eigenvalues, which are complex, according
to whether they are continuations
from respulsive or attractive eigenvalues at negative energy.} 
We note that  lowering the cutoff 
and increasing the density both contribute to 
dampening the sources of nonperturbative physics, which is
reflected by the  smaller magnitudes of the eigenvalues at finite
densities, consistent with the nuclear  matter results of
Ref.~\cite{perturbative}. 

Close to the Fermi surface, the attractive interaction between the 
particles leads to the normal ground state becoming
unstable to the formation of Cooper pairs.  
While considering the behavior of the eigenvalues away from the Fermi 
surface (e.g., for analyzing bulk properties),
it is sufficient to consider only the particle-particle Green's function, 
because the
phase space for hole propagation is small~\cite{FetterWalecka}.
Investigating the signatures of pairing, however,
requires us to consider the hole
propagation as well, since the phase space around the Fermi surface becomes
important.     

\bfig[t]
\begin{center}
	\includegraphics*[angle=0,width=5in]
{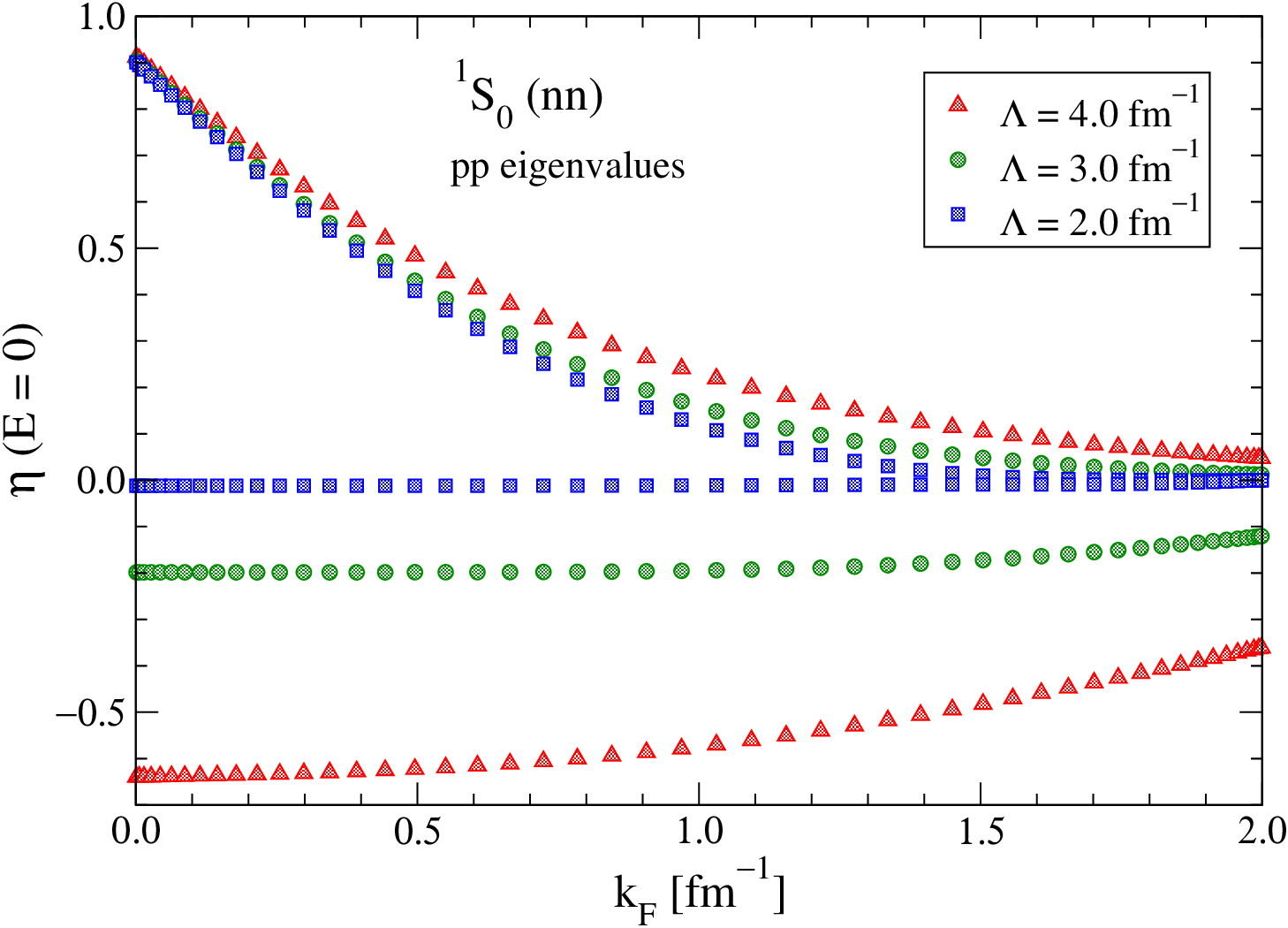}
\vspace*{-.1in}
  \caption{Largest attractive ($\eta > 0$)
  and repulsive ($\eta < 0$) eigenvalues at $\Ecm = 0 \mev$ for
  neutron matter in the $^1S_0$ channel for three different cutoffs,
  as a function of Fermi momentum $\kf$.}  
  \label{med.fig0}
\end{center}
\efig	  
  
The in-medium Weinberg eigenvalue equation including the hole propagation is:
\beq
    \Gmed(E) V |\Psi_{\nu}(E)\ran = \eta_{\nu}(E)|\Psi_{\nu}(E)\ran \;,
    \label{med.eqn1}
\eeq		
where $\Gmed(E)$ is the in-medium two-particle and two-hole non-interacting 
propagator. In momentum space, this propagator is given 
by~\cite{FetterWalecka}:
\beq
    \Gmed(\veck_1, \veck_2;\omega) = \D\frac{\theta(|\veck_1| - \kf)\theta
    (|\veck_2| - \kf)}{\omega - \varepsilon(\veck_1) - \varepsilon(\veck_2)
     + i\epsilon} - \D\frac{\theta(\kf - |\veck_1|)\theta(\kf - |\veck_2|)}
     {\omega - \varepsilon(\veck_1) - \varepsilon(\veck_2) - i\epsilon}
   \;, \label{med.eqn2}
\eeq	
where $\veck_1$ and $\veck_2$ are the momenta of the particles (holes),
$\kf$ is the Fermi momentum,
and 
$\omega$ is the two-particle excitation energy measured from the Fermi surface. 
The above propagator represents propagation of two particles above the Fermi 
surface and two holes below the Fermi surface. We use the free-particle
spectrum to be consistent with the usual BCS treatment applied in 
Refs.~\cite{HebelerSchwenk:2006}, \cite{Lombardo:2000ec}, and \cite{SchwenkPairing1}. 

The phase space for pairing is maximal for back-to-back
pairs~\cite{schrieffer},
therefore we apply Eq.~(\ref{med.eqn2}) with zero center-of-mass
momentum,
\beq
	 \Gmed(k,\omega) = \D\frac{\theta(k - \kf)}{\omega - 2\varepsilon(k)  
	 + i\epsilon} - \D\frac{\theta(\kf - k)}{\omega - 2\varepsilon(k) - i
	 \epsilon} \;,
	\label{med.eqn3}
\eeq 
where $\vec{k}$ is the relative
momentum.  With
$\omega = E - 2\mu$ and 
$\varepsilon(k) = k^2/2 - \mu$, where $\mu = \kf^2/2$ is  
the zero-temperature, non-interacting Fermi energy
(we use units in which 
${\hbar^2}/{m_N} = 1$, with $m_N$ the mass of a nucleon), 
the in-medium propagator is simply
\beq
    \Gmed(k, E = k_0^2) = \D\frac{\theta(k - \kf)}{k_0^2 - k^2  + 
   i\epsilon} - \D\frac{\theta(\kf - k)}{k_0^2 - k^2 - i\epsilon}
   \;.
   \label{med.eqn4}
\eeq 

We now study the Weinberg eigenvalues for the kernel $\Gmed(E) \vlowk$. 
Just as for the free-space case~\cite{Bogner:2006tw}, we actually solve
for the eigenvalues of 
$\vlowk \Gmed(E)$, which has the same eigenvalue spectrum but
allows for direct integration over singularities. 
In a given partial wave, the Weinberg eigenvalue equation is:
\beq 
    \D\frac{2}{\pi} \int^{\la}_0 q^2 dq \,
    \vlowk(k,q) \left( \D\frac{
    \theta(q - \kf)}{k_0^2 - q^2  + i\epsilon} - \D\frac{\theta(\kf - q)}
    {k_0^2 - q^2 - i\epsilon} \right) \Psi_{\nu}(q) = \eta_{\nu}(k_0^2) 
    \Psi_{\nu}(k) \;.
   \label{med.eqn5}
\eeq
For notational convenience
we have suppressed the energy dependence of the eigenvectors
in Eq.~(\ref{med.eqn5}). Using 
the standard identities,
\beq
     \D\frac{1}{x - x_0 \pm i\epsilon}= {\cal P} \D\frac{1}{x - x_0} 
    \mp i\pi\delta(x  - x_0)  \;, 
    \label{med.eqn6}
\eeq
\beq	
     \delta(f(x))=\sum_i \D\frac{\delta(x - 
    x_i)}{|f^{\prime}(x)|_{x = x_i}}  \;, 
    \label{med.eqn7}
\eeq
Eq.~(\ref{med.eqn5}) becomes:
\bea
     \lefteqn{\D\frac{2}{\pi} {\cal P} \int^{\la}_0 q^2 dq\, \vlowk(k,q)  
    \left( \D\frac{\theta(q - \kf) - \theta(\kf - q)}{k_0^2 - q^2} \right) 
    \Psi_{\nu}(q)} \nonumber \\
    && \hspace*{1.5in}\mbox{} - i k_0 \vlowk(k,k_0)  \Psi_{\nu}(k_0) = 
    \eta_{\nu}(k_0^2) \Psi_{\nu}(k)  \;.
    \label{med.eqn8}
\eea	

\bfig[t]
\begin{center}
	\includegraphics*[angle=0,width=5in]
{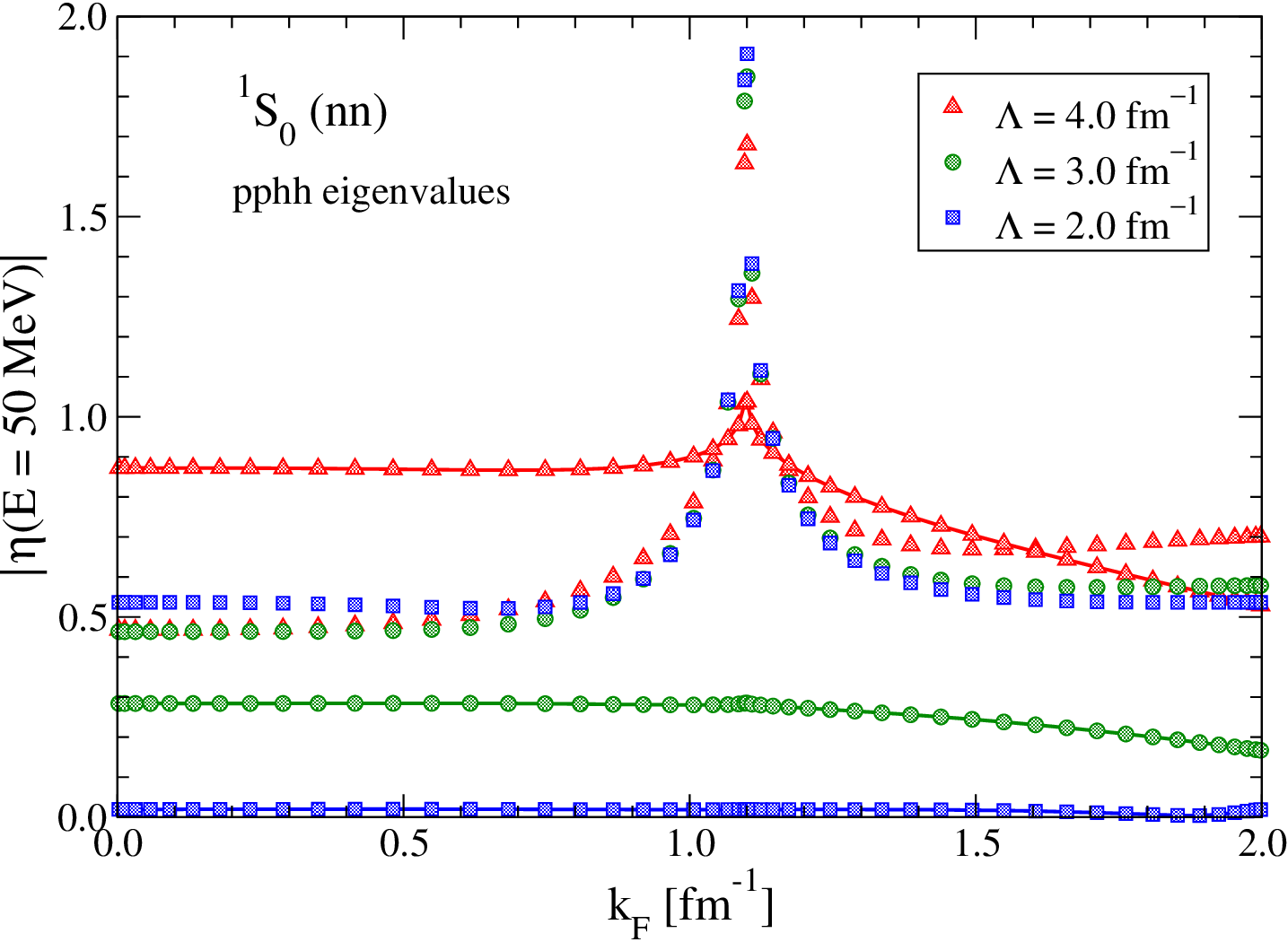}
\vspace*{-.1in}
  \caption{Magnitudes of the
  largest attractive and repulsive eigenvalues at $E = 50 \mev$ 
  as a function of Fermi momentum $\kf$ for
  neutron matter in the $^1S_0$ channel for three different cutoffs
  using the two-particle and two-hole propagator. 
  (The repulsive eigenvalues are connected by lines.)} 
  \label{med.fig0a}
\end{center}
\efig	  

\bfig[t]
\begin{center}
	\includegraphics*[angle=0,width=5in]
{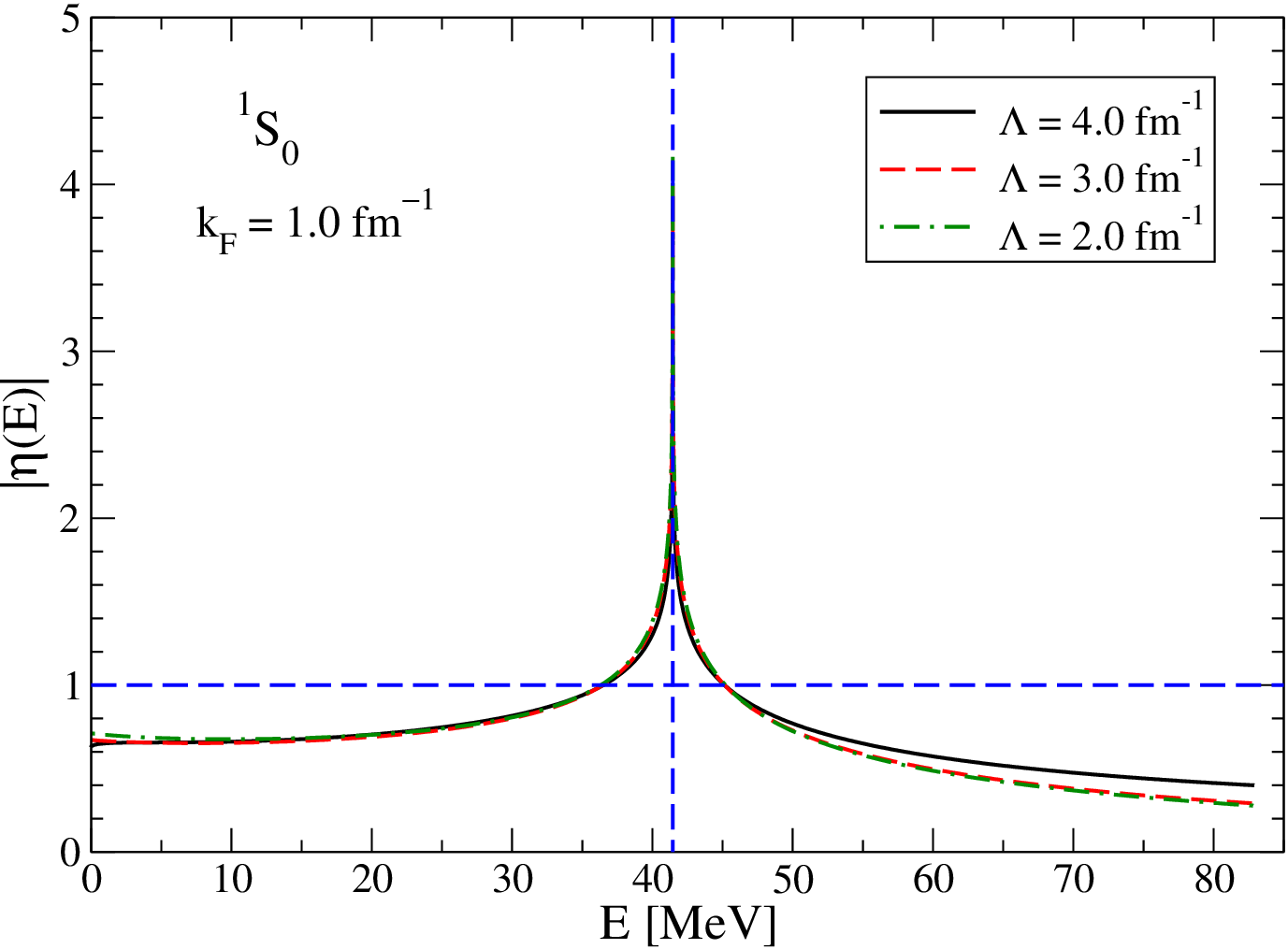}
\vspace*{-.1in}
  \caption{ Magnitude of the largest attractive eigenvalue for neutron matter 
  as a function of energy
  $E$  in the $^1S_0$ channel for three different cutoffs
  using the two-particle and two-hole propagator.}  
  \label{med.fig1}
\end{center}
\efig	  

To identify signatures of pairing,
we plot in Fig.~\ref{med.fig0a} the magnitude of the largest
attractive and repulsive Weinberg eigenvalues for $G^0_{pphh}V$
as a function of density    for the $^1S_0$ partial wave
with center-of-mass energy 
$E = 50\,\mev$.
The $\vlowk$ matrix elements are from the Argonne $v_{18}$
potential~\cite{AV18} using a sharp regulator, but the results here
apply generally
to all low-momentum potentials.
As we scan through $\kf$, the largest repulsive eigenvalue for larger
cutoffs shows a cusp behavior due to the sharp
Fermi surface that is localized near 
the momentum corresponding to $\kf = \sqrt{E}$. 
The repulsive eigenvalues are strongly cutoff dependent and no
cusp is resolved at the lower cutoffs.
In contrast, the attractive eigenvalues depend weakly on cutoff and
show a broad cusp near the Fermi surface; this behavior is the
link to the pairing instablity. 
Similarly,
Fig.~\ref{med.fig1} shows the largest attractive eigenvalue at a
fixed density (in this case $\kf = 1.0 \fmi$) as a function of 
$E$ at several cutoffs. 
Again we find weak cutoff dependence and a broad cusp about the
Fermi surface; in the next section we extract the
gap from this behavior by going to complex energies.


\subsection{Stability Analysis and Pairing gaps}
\label{sect:stability}

In this section we adapt the stability analysis of Ref.~\cite{schrieffer}.
Consider Eq.~(\ref{med.eqn4}) for the two-particle and two-hole
non-interacting Green's function. Above $2\ef$ we have the particle-particle
continuum and below $2\ef$ we have the hole-hole continuum. Therefore, a
stable
bound state around $2\ef$ cannot be accomodated. In fact, 
the bound-state energies measured from the Fermi
surface $2\ef$
are purely imaginary  and have a value of $\pm i
\Delta_F$, where $\Delta_F$ is the BCS gap at 
$\kf$~\cite{schrieffer,DickhoffManybody} (actually just an approximation, 
see below).
This result can be easily established by studying the singularity
structure of the in-medium $T$ matrix in the complex 
plane~\cite{schrieffer}. Two purely
imaginary poles of the $T$ matrix appear for an attractive two-body
interaction between the pairs and the  normal phase becomes unstable.  
Therefore, at $E = 2\ef \pm i \Delta_F$ the magnitude of the largest
attractive eigenvalue $|\eta(2 \ef \pm i \Delta_F)|$ equals one,
signaling the presence of a bound state at these energies. Dialing the
imaginary part of $2 \ef \pm i E_0$ from negative energies through $0$
to  positive energies leads to eigenvalues that start to grow as
$E_0$  increases, cross one at $E_0 = \pm \Delta_F$ and become singular
as $E_0 \rightarrow 0$. The eigenvalues are symmetric about $E_0 = 0$.

\bfig[p]
\begin{center}
	\includegraphics*[angle=0,width=4.9in]
{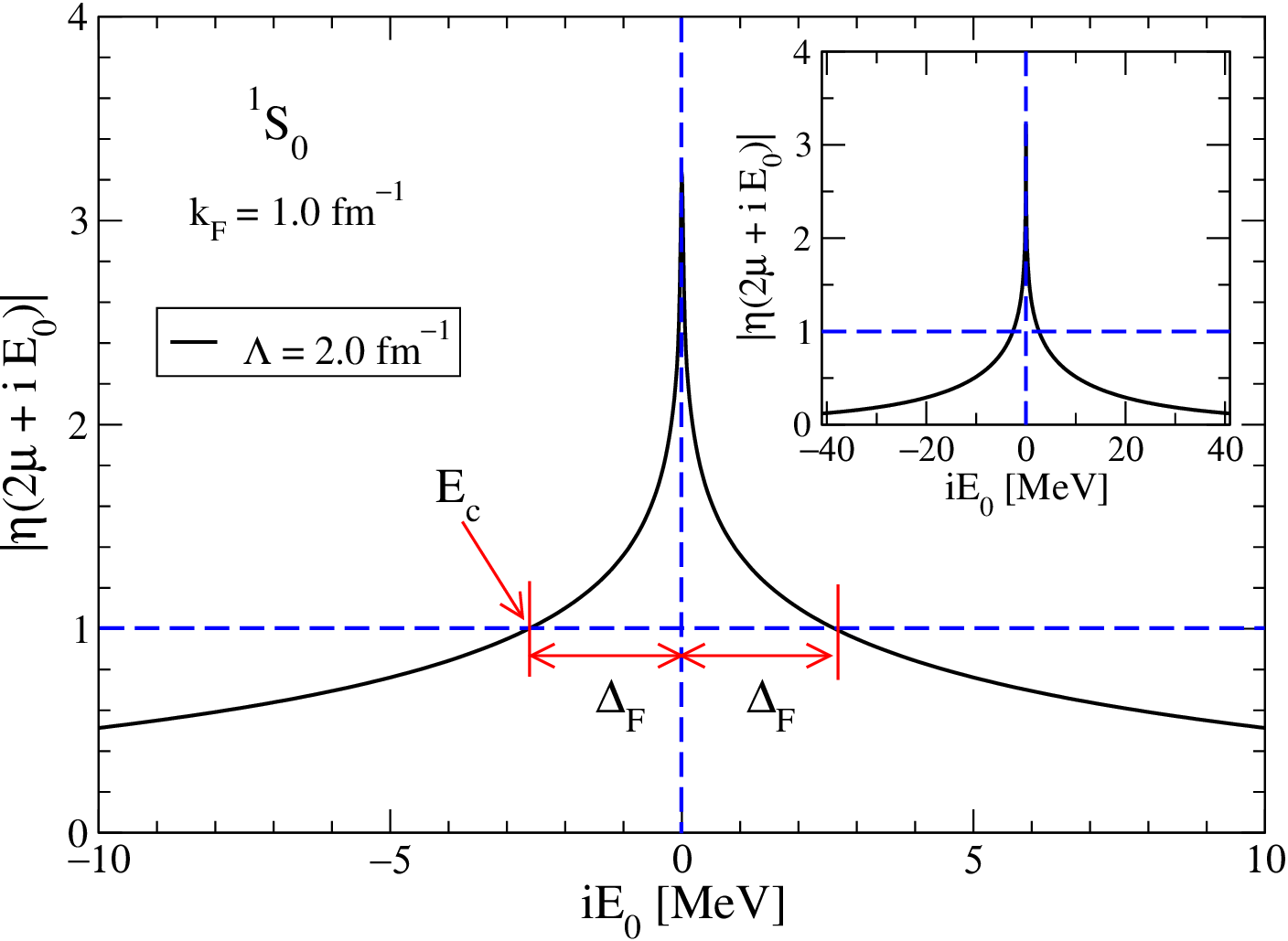}
\vspace*{-.1in}
	\caption{Largest attractive eigenvalue as a function of energy
	$iE_0$ (scanning along the imaginary axis). 
        $E_c$ refers to the critical energy for which $|\eta_{\nu}| = 1$;
        in this
	case $E_c \approx \Delta_F$, the BCS gap at $\kf$.}  
	\label{instab_1s0.fig}
\end{center}
%
\vspace*{.15in}
\begin{center}
        \includegraphics*[angle=0,width=4.9in]{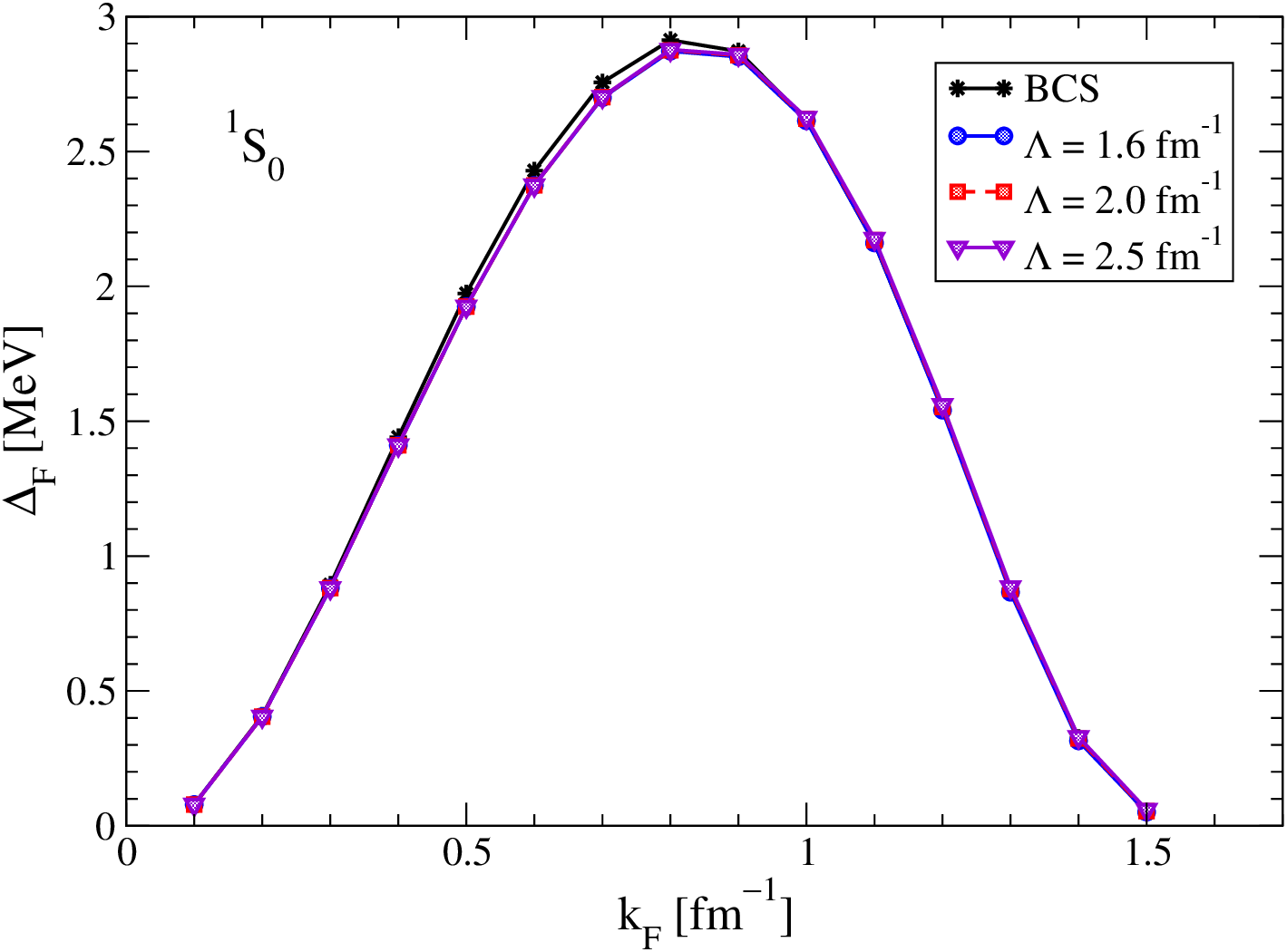}
	\vspace*{-.1in}
	\caption{Density dependence of the gap extracted through the
        stability analysis for neutron 
	matter at several different
        cutoffs $\la$ compared to the self-consistent gaps obtained from the BCS gap
	equation.}  
	\label{gap_vs_kf.fig}
\end{center}
\efig

This behavior is seen in Fig.~\ref{instab_1s0.fig}, where we plot the largest 
attractive Weinberg eigenvalue in the $^1S_0$ channel for neutron matter at
$\kf = 1.0 \fmi$ for a cutoff of $\la = 2.0 \fmi$. 
The value of the imaginary part of the energy 
where the eigenvalue crosses one ($E_c$) directly gives 
a first approximation to the pairing gap.  
In Fig.~\ref{gap_vs_kf.fig} we show the corresponding density dependence 
of the pairing gaps extracted via the stability analysis for several
cutoffs 
compared to self-consistent gaps obtained from the BCS gap
equation~\cite{HebelerSchwenk:2006}. 
Note that we 
are working in the limit where $\Delta_F/\ef$ is small, so that the gap 
is independent of the momentum $k$ and depends only on the density $\kf$. 
The errors in the gaps obtained from the
stability analysis scale as a power of ${\Delta_F}/{\ef}$. 
Thus the in-medium Weinberg eigenvalues not only reflect the
instability  of the normal phase, they also give a good estimate of the
pairing gap.  
For the range of cutoffs
($\la = 1.6 \fmi$ to $2.5 \fmi$) considered here, Fig.~\ref{gap_vs_kf.fig}
shows that the gaps exhibit very weak cutoff
dependence, as was found in
Ref.~\cite{HebelerSchwenk:2006}.

The gap can be cleanly
extracted from the stability analysis at lower cutoffs because the
effect of other sources of non-perturbative physics has been dampened,
leaving the largest attractive eigenvalue isolated.  The analysis is
more involved at higher cutoffs, but still possible in the $^1$S$_0$
channel, where repulsive eigenvalues corresponding to the strong
short-range repulsion dominate.  By continuing the attractive
eigenvalue from zero energy, where it is cleanly distinguished, to
the Fermi surface, gaps can be extracted at all cutoffs.  As in
Fig.~\ref{gap_vs_kf.fig}, these agree well with the BCS gaps and
show only very small cutoff dependence at all higher cutoffs.


\section{Weinberg Eigenvalues in the Nambu-Gorkov Formalism}
  \label{sect:nambu}

Thus far we have worked in the normal phase with the non-interacting
propagator.  Here we consider the Weinberg eigenvalue analysis in 
the paired phase using the Nambu-Gorkov formalism.
In momentum space for a homogeneous system, the corresponding Nambu-Gorkov
Green's function reduces to~\cite{AbrikosovGorkov} 
\beq
	 G^{0}_{\rm NG}(\veck, \omega) = \int \D\frac{d\omega^{\prime}}{2\pi i} 
	 \left[ \Gprop(\veck,(\omega))\Gtildeprop(\veck,\omega - 
	 \omega^{\prime}) + \Fprop(\veck, \omega)\Fdagprop(\veck, 
	 \omega - \omega^{\prime}) \right] \;, 
	\label{med.eqn27}
\eeq
where the normal and anomalous propagators are given 
by
\beq
	 \Gprop(\veck, \omega) = \D \frac{u_k^2}{\omega - E_k + i\epsilon} + 
	 \frac{v_k^2}{\omega + E_k - i\epsilon}  \;,	
	\label{med.eqn21}
\eeq
\beq
	 \Gtildeprop(\veck, \omega) = - \D \frac{u_k^2}{\omega + E_k - 
	 i\epsilon} - \frac{v_k^2}{\omega - E_k + i\epsilon}  \;,	
	\label{med.eqn22}
\eeq
\beq
	 \Fprop(\veck, \omega) = \Fdagprop(\veck, \omega) = - u_k v_k 
\left( \D \frac{1}{\omega - E_k + i\epsilon} - \frac{1}{\omega + E_k - 
i\epsilon} \right)	\;.
	\label{med.eqn23}
\eeq
The spectral functions $u_k$ and $v_k$ are defined as
\beq
	 u_k^2  = \D \frac{1}{2}\left(1 + \frac{\xi_k}{E_k} \right) 
	 \;, \qquad
	 v_k^2  = \D \frac{1}{2}\left(1 - \frac{\xi_k}{E_k} \right)
         \;, 
	 \label{med.eqn25}
\eeq
with 
\beq
	 E_k  =  \sqrt{\xi_k^2 + \Delta(k)^2}  \;. 
	\label{med.eqn26}
\eeq
Here $\xi_k = \ek - \mu$ is the single-particle energy measured from 
the chemical potential $\mu$, which 
for a non-interacting system at zero temperature is $\mu = {\kf^2}/{2}$. 
In Eq.~(\ref{med.eqn26}), $\Delta(k)$ is  the gap function.  
Evaluating the contour integral in Eq.~(\ref{med.eqn27}) we get the 
following expression for the two-particle Nambu-Gorkov propagator,
\beq
	 G^{0}_{\rm NG}(\veck, \omega) = \D \frac{u_k^2}{\omega - 2E_k + i\epsilon} - 
	 \frac{v_k^2}{\omega + 2E_k - i\epsilon} \;.
	\label{med.eqn28}
\eeq
Taking the $\Delta(k) \rightarrow 0$ limit reproduces the non-interacting 
two-particle $\Gmed$ propagator.

The Weinberg eigenvalue equation is (with $E = \omega + 2\mu$)
\beq
    G^{0}_{\rm NG}(k, E)V|\Psi_{\nu}(E)\ran  = \etaNG(E) |\Psi_{\nu}(E)
    \ran  \;.
    \label{eq:genNG} 
\eeq
At $E = 2\mu$  the 
eigenvalue problem can be reduced to the BCS gap equation. To 
see this, we start with Eq.~(\ref{eq:genNG}) for
a general potential in momentum
space,
\beq
	 \D\frac{2}{\pi} \int_0^{\la} k^2 dk\,
          G^{0}_{\rm NG}(\kprime, E) 
	 V(\kprime, k)
	 \Psi_{\nu}(k, E) = \etaNG(E)\Psi_{\nu}(\kprime, E) \;. 
	\label{med.eqn37}
\eeq	
As noted earlier, we can solve the above equation by converting it to a 
left-eigenvalue problem, or alternatively solve the right-eigenvalue problem,
\beq
	 VG^{0}_{\rm NG}(z)[V|\Psi_{\nu}(z) \ran] = \etaNG(z) [V|\Psi_{\nu}(z) 
	 \ran]  \;, 
	\label{med.eqn38}
\eeq	
which has the same spectrum as $G^{0}_{\rm NG}(z)V$, and integrate over the 
singularity 
directly. 
Using the two-particle 
Nambu-Gorkov propagator, 
Eq.~(\ref{med.eqn38}) 
can be written as
\bea
	 \lefteqn{\D\frac{2}{\pi} \int_0^{\la} k^2 dk V(\kprime, k) 
	 \left(\D \frac{u_k^2}{\omega - 2E_k + i\epsilon} - \frac{v_k^2}
{\omega + 2E_k - i\epsilon} \right) \Psi_{\nu}(k, E)}\nonumber \\
	&& \hspace*{2.2in} \mbox{} = \etaNG(E)\Psi_{\nu}(\kprime, E)
	\;. 
	\label{med.eqn39}
\eea	
For $\omega = 0$ (i.e., $E=2\mu$) and with
$E_k = \sqrt{\xi_k^2 + \Delta(k)^2}$, Eq.~(\ref{med.eqn39}) has
a solution with $\etaNG(2\mu) = 1$:
\beq
	 - \D\frac{1}{\pi}\int_0^{\la} k^2 dk V(\kprime, k) 
	 \left(\D \frac{1}{\sqrt{\xi_k^2 + \Delta(k)^2}}\right) \Psi_{\nu}
	 (k, 2\mu) 
	 = \Psi_{\nu}(\kprime, 2\mu)  \;, 
	\label{med.eqn40}
\eeq	
which is equivalent to the gap equation with
the corresponding 
eigenvector $\Psi_{\nu}(k, E \rightarrow 2\ef)$ equal to the gap function 
$\Delta(k)$.
However,
it is not clear how to 
derive $\Delta(k)$ from 
the eigenvalue equation because of the dependence of $E_k$ on $\Delta(k)$;
simple iterations do not lead to self-consistency.
At energies of $2\ef \pm 2\Delta_F$, the eigenvalue 
in Eq.~(\ref{med.eqn39}) becomes singular, 
see Fig.~\ref{med.fig6}. This coincides with the branch point of the 
two-particle Green's function and can be associated with 
the breaking of the pairs.

\bfig[t]
\begin{center}
  \includegraphics*[angle=0,width=5in]{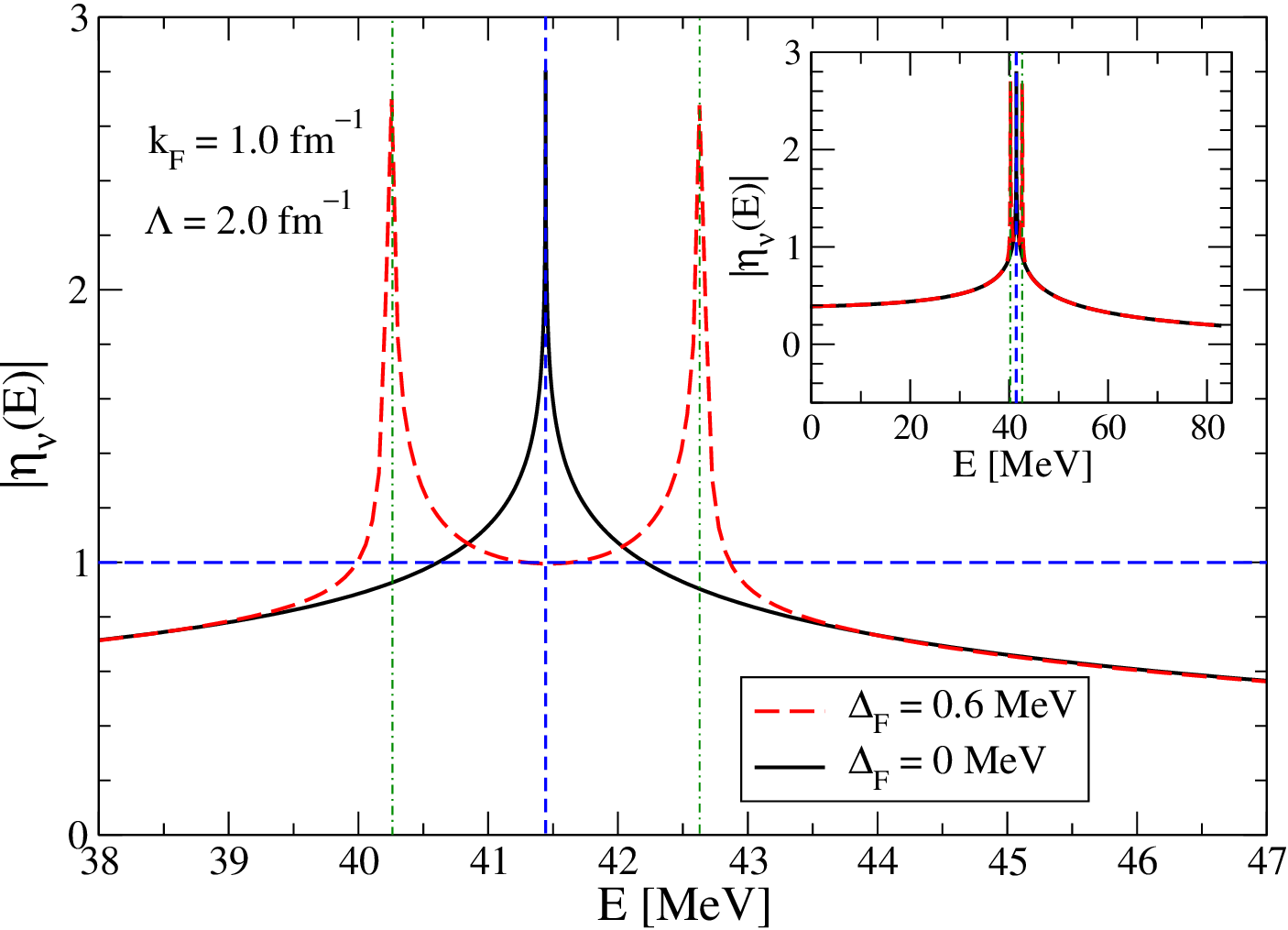}
  \caption{Plot of the Weinberg eigenvalue for the separable potential of
  Eqs.~(\ref{med.eqn29}) and (\ref{med.eqn35}) as a function of $E$ for 
  both $\Delta_F = 0 \mev$ and $\Delta_F \approx 0.604 \mev$. The solid 
  line represents the eigenvalue for the non-interacting propagator,
  $\Gmed$,  while the dashed lines is for the Nambu-Gorkov propagator,
  $G^{0}_{\rm NG}$.  The Nambu-Gorkov  eigenvalue $\etaNG$ goes
  to one close to $2\ef$, indicating  formation of a Cooper pair. Large
  singular values for  $\etaNG$ are observed at $2\ef \pm 2\Delta_F$,
  which represents the energy at  which the pairs are broken.}  
	\label{med.fig6}
\end{center}
\efig

To illustrate the behavior near the
Fermi surface,
we first calculate the Weinberg eigenvalues of the Nambu-Gorkov 
propagator using a separable potential of the form
\beq
	 V = \lam |f \ran \lan f|  \;,
	\label{med.eqn29}
\eeq		
where $\lam$ is a coupling that controls the strength of the potential. $\lam$
is negative for an attractive potential and positive for a repulsive potential. 
The Weinberg eigenvalue equation in momentum space is
\beq
   \D\frac{2}{\pi}\lam \lan f|\Psi_{\nu}(E)\ran \int_0^{\la} k^2 dk 
   G^{0}_{\rm NG}(k, E) \lan k|f\ran |k\ran  = \etaNG(E)|\Psi_{\nu}(E)
   \ran  \;. 
   \label{med.eqn31}
\eeq
From Eq.~(\ref{med.eqn31}), we see that there is only one 
Weinberg eigenvalue for any rank-one separable potential, which is given by
\beq
   \etaNG(E) = \lam \lan f|\Psi_{\nu}(E)\ran  \;,
    \label{med.eqn32}
\eeq
and the corresponding eigenvector is
\beq
    |\Psi_{\nu}(E)\ran = \D\frac{2}{\pi} \int_0^{\la} k^2 dk \,
    G^{0}_{\rm NG}(k,E)f(k) |k\ran  \;.
   \label{med.eqn33}
\eeq
Substituting Eq.~(\ref{med.eqn33}) into 
Eq.~(\ref{med.eqn32}), we get the following closed expression for 
the Weinberg eigenvalue:
\beq
	 \etaNG(E) = \lam \D\frac{2}{\pi} \int_0^{\la} k^2 dk 
	 G^{0}_{\rm NG}(k,E)|f(k)|^2  \;.
	\label{med.eqn34}
\eeq					

For simplicity we choose 
the function $f(k)$ to be a gaussian, 
\beq
	 f(k) = e^{-k^2/\alpha^2} \;,
	\label{med.eqn35}
\eeq
with $\alpha = \sqrt{2} \fmi$.  We also set $\lam = - 1\,{\rm fm}$, so 
that the separable potential is attractive.
Figure~\ref{med.fig6} shows the Weinberg eigenvalue for the 
separable 
gaussian potential as a function of center-of-mass energy $E$ for 
$\Delta_F = 0 \mev$ and for finite $\Delta_F$. 

For vanishing $\Delta_F$, the eigenvalue becomes singular at $E = 2\ef$ due to
the sharp Fermi surface. For the gapped phase, the gap function appearing
in the NG propagators can be estimated from the normal-phase Weinberg
analysis for which $\eta_{\nu}(2\ef + i\Delta_F) = 1$. 
In the present separable
example, the normal-phase estimate gives $\Delta_F \approx 0.604 \mev$. 
This value is
taken as the initial guess for the gap function, $\Delta(k) \approx
\Delta_F$, which
is then used in the two-particle Nambu-Gorkov propagator
and the corresponding eigenvalue is 
evaluated using
\beq
    \etaNG(E) = \lam \D\frac{2}{\pi}  \int_0^{\la} k^2 dk 
    |f(k)|^2 
    \left(\D \frac{u_k^2}{\omega - 2E_k + i\epsilon} - \frac{v_k^2}
   {\omega + 2E_k - i\epsilon} 
	 \right)  \;,
	\label{med.eqn36}
\eeq
where $\omega = E - 2 \mu$, and $E_k = \sqrt{\xi_k^2 + \Delta_F^2}$. 
We see that the  eigenvalue  $\etaNG(E)$ tends to one
close to $E = 2\mu$, as expected. 
 
\bfig[t]
\begin{center}
	\includegraphics*[angle=0,width=5in]
{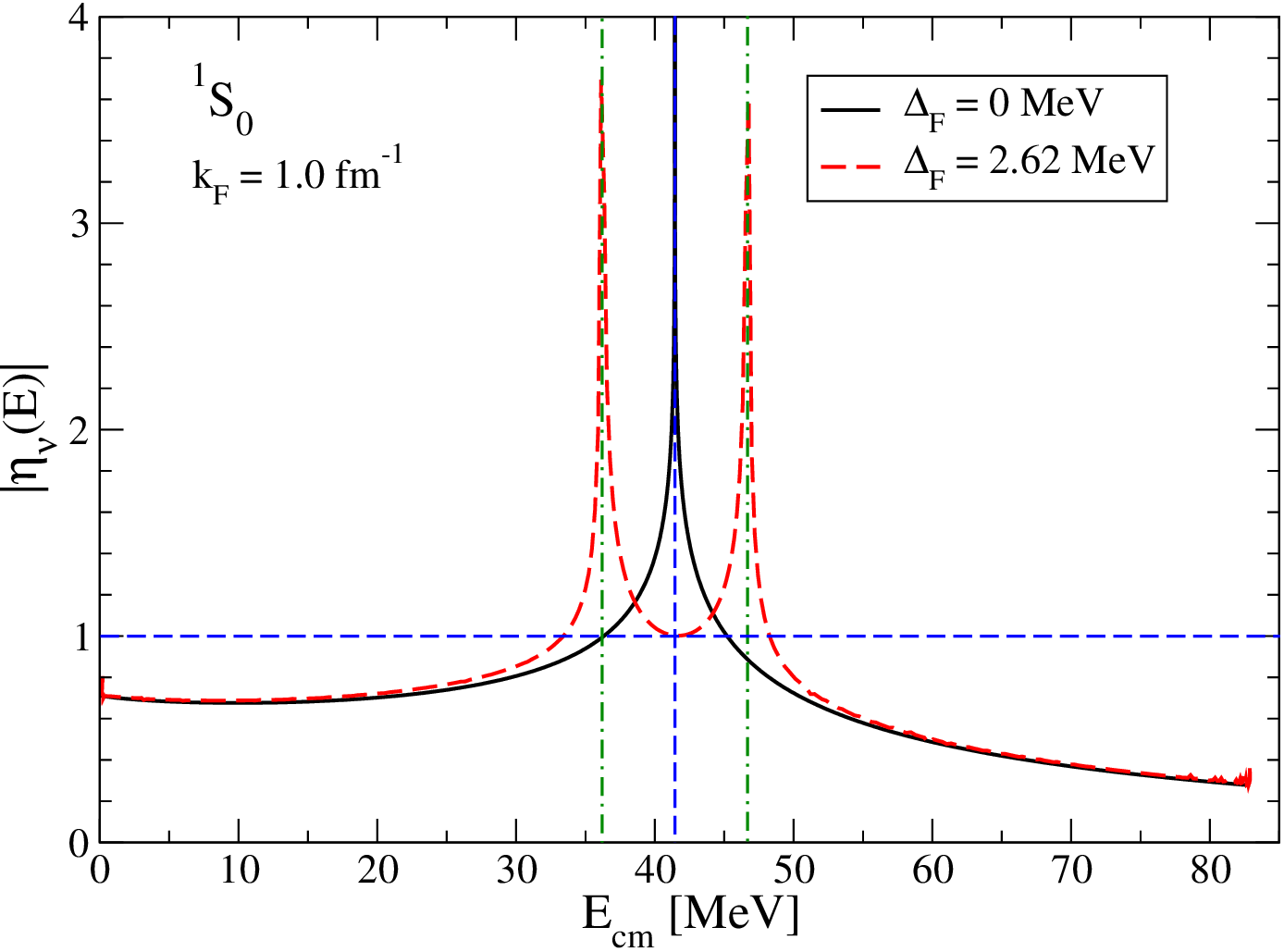}
	\caption{Largest attractive Weinberg eigenvalue as a function 
	of energy $E$ for a cutoff $\la = 2.0 \fmi$ at $\kf = 1.0 \fmi$. 
        Notice that as $E \rightarrow 2\ef$, 
	$\eta_{\nu}(E)$ tend to one.}
	\label{med.fig8}
\end{center}
\efig

\bfig[p]
\begin{center}
	\includegraphics*[angle=0,width=3.5in]
{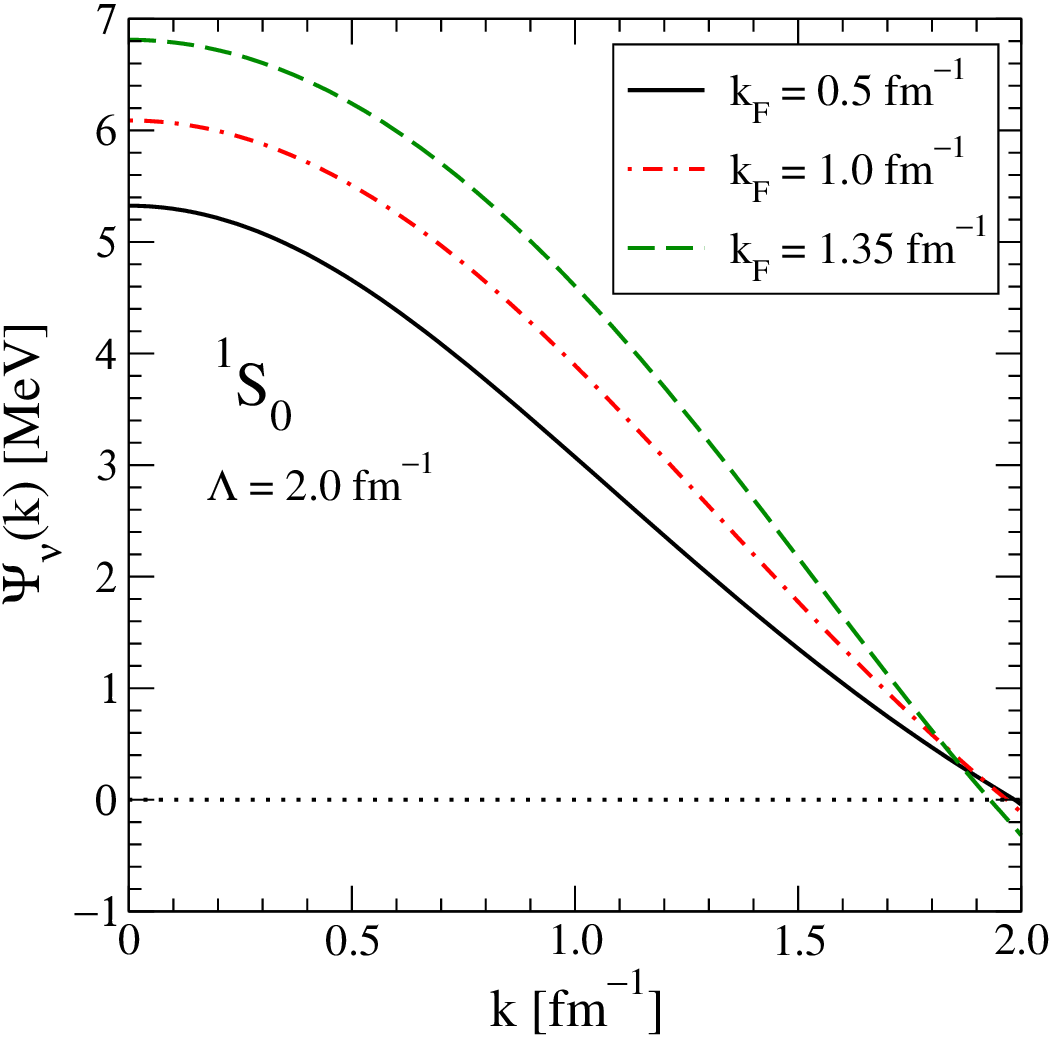}
	\caption{Momentum dependence of the eigenvector corresponding to the 
	largest attractive eigenvalue at  $\omega = 0$ for representative $\kf$ values. This serves as the first
	approximation to the gap function.}  
	\label{med.fig9}
\end{center}
%
\vspace*{.15in}
\begin{center}
	\includegraphics*[angle=0,width=3.5in]
{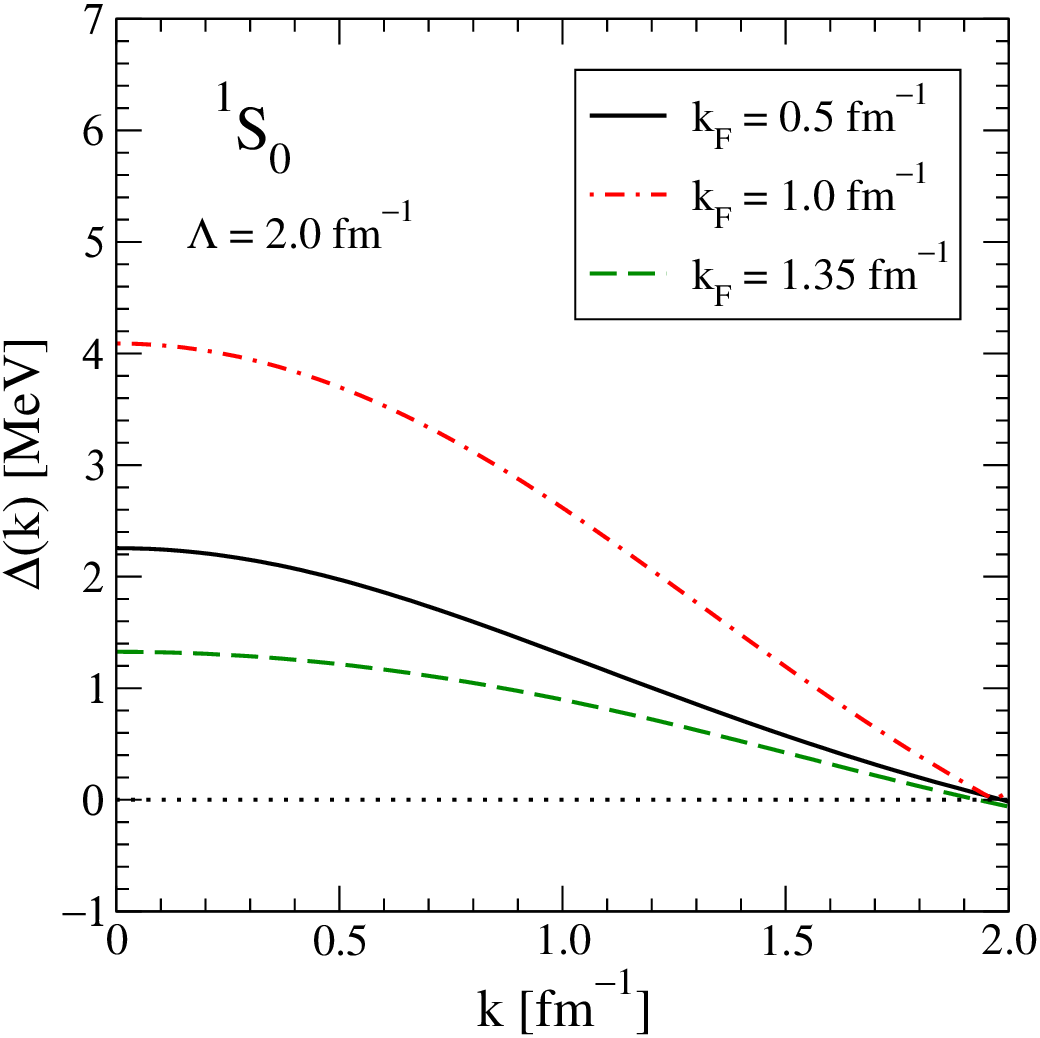}
	\caption{Momentum dependence of the self-consistent gap function 
	$\Delta(k)$, obtained from the Weinberg eigenvectors corresponding 
	to the largest attractive eigenvalue at $\omega = 0$
	as the initial gap function.}  
	\label{med.fig10}
\end{center}
\efig	  

Figure~\ref{med.fig8} shows the largest attractive Weinberg 
eigenvalue as a function of $E$ for $\vlowk$ in the $^1S_0$ partial wave 
at $\kf = 1.0 \fmi$ for a cutoff $\la = 2.0 \fmi$. 
As with the separable example, the gap function $\Delta(k)$ appearing in the
NG propagators is taken from the normal-phase Weinberg eigenvalue
estimate.
Apart from some small
numerical instabilities outside the 
$\pm 2\Delta_F$ region, the largest eigenvalue behaves like the 
eigenvalue in the separable case  as 
$E \rightarrow 2\ef$  (see Fig.~\ref{med.fig6}). 
Similarly, the eigenvector corresponding to $\etaNG(E) = 1$ 
as $E \rightarrow 2\mu$ is an approximation to 
the gap function. 
Figure~\ref{med.fig9} shows 
the eigenvector corresponding to the largest eigenvalue at $E = 2 \ef$ 
for representative densities. 
Using this as the first guess for the gap function $\Delta(k)$ and iterating 
the gap equation,
\beq
	\Delta(k) = -\D\frac{1}{\pi} \int_0^{\infty} q^2 dq 
	\frac{V(k, q) \Delta(q)}{\sqrt{\xi_q^2 + \Delta(q)^2}} \;,
	\label{med.eqn48}
\eeq		
yields the self-consistent BCS gap 
function, which is shown in Fig.~\ref{med.fig10}. 
It is evident that the non-self-consistent eigenvector is a poor
approximation to $\Delta(k)$ in general.
As $\kf$ increases, the gap 
closes for smaller momentum values, as observed in 
Ref.~\cite{KhodelKhodel}. These results are consistent with the 
density dependence of the $^1S_0$ gap $\Delta_F$.

\section{Summary}  
\label{sect:conclusion}

In summary, we see that pairing instability is reflected in the behavior of the
Weinberg eigenvalues close to the Fermi surface, with the signature of
nonperturbativeness given by the presence of eigenvalues outside the
unit circle. We find that a good approximation to the momentum independent gaps
$\Delta_F$ can be obtained from a stability analysis, which means that  
weak coupling is a good approximation for low-momentum interactions.
If instead we use the
Nambu-Gorkov  two-particle Green's function, the largest attractive eigenvalue
tends to one close to the Fermi surface, indicating the presence of bound
states (Cooper pairs).   At the Fermi surface, the eigenvalue equation is the
gap equation and the self-consistent
eigenvector corresponding to the largest attractive
eigenvalue is the gap function.  

The gaps we have shown correspond to the BCS results and do not include 
polarization effects. These effects are known to significantly 
reduce the gap in neutron 
matter~\cite{SchwenkPairing1,HeiselbergPethick:2000,Lombardo:2000ec,SchwenkPairing:2004,SchwenkPairing:2006}. 
These interactions can be taken into account within this
framework using the potential
\beq
   V(\kprime, k) = V_0(\kprime, k) + V_{\rm ind}(\kprime, k),
  \label{med.eqn53}
\eeq
where $V_0(\kprime, k)$ is the two-body interaction and 
$V_{\rm ind}(\kprime, k)$ is the induced interaction, given in detail 
in Refs.~\cite{HeiselbergPethick:2000,schulze1996}. Once again we 
use the two-particle Green's function, $\Gmed$, to calculate the eigenvalues 
close to the Fermi-surface and determine the pairing gap at $\kf$ in a 
similar fashion. This method therefore offers an extension to 
include medium effects. 

 Any residual cutoff dependence in the gaps suggests the importance of
many-body forces. 
Our results agree with the
recent work of Hebeler et al~\cite{HebelerSchwenk:2006} 
that shows that
the cutoff dependence of the $^1S_0$ gap in neutron matter is weak. As a result
the three-nucleon contribution to the $^1S_0$ pairing gaps in neutron matter is
expected to be
small at the BCS level. A preliminary analysis for the $^3S_1$ gaps in nuclear
matter  exhibit strong cutoff dependences and hence it would be worthwhile   to
investigate the role of three-nucleon interactions and the corresponding medium
effects.

The success of the in-medium Weinberg analysis for pairing suggests that it
should be useful in assessing other possible sources of nonperturbative
physics.  Recent results by Roth and collaborators~\cite{rothRPA}  indicate
that, for low-momentum potentials,   correlation effects in finite nuclei
beyond second order do not significantly change the binding  energy per
nucleon. This motivates us to investigate the perturbativeness of the
particle-hole channel for bulk nuclear matter using the Weinberg
eigenvalue analysis, which is in progress. 

\begin{ack}
We thank Achim Schwenk for useful comments and discussions. This work was 
 supported in part by the National Science Foundation
under Grant Nos.~PHY--0354916 and PHY--0653312  and by the U.S. Department 
of Energy under 
grant DE-FG02-93ER40756.
\end{ack}


\end{document}